\documentclass{article}

\usepackage{amssymb}
\usepackage{amsmath}
\usepackage{graphicx}

\begin{document}


\title{Managing catastrophic changes in a collective} \label{chap:csmag}

\author{David Lamper, Paul Jefferies, Michael Hart and Neil F. Johnson\\
Oxford University, Parks Road, Oxford OX1 3PU, U.K.}

\maketitle

\begin{abstract}
We address the important practical issue of understanding, predicting and
eventually controlling catastrophic endogenous changes in a collective. Such
large internal changes arise as macroscopic manifestations of the
microscopic dynamics, and their presence can be regarded as one of the
defining features of an evolving complex system. We consider the specific
case of a multi-agent system related to the El Farol bar model, and show
explicitly how the information concerning such large macroscopic changes
becomes encoded in the microscopic dynamics. Our findings suggest that these
large endogenous changes can be avoided either by pre-design of the
collective machinery itself, or in the post-design stage via continual
monitoring and occasional `vaccinations'.

\vspace{5mm}
\noindent
A contribution to the Workshop on Collectives and the Design of Complex
Systems, organized by David Wolpert and Kagan Tumer, at NASA Ames Research
Center, CA, August (2002).

\end{abstract}

\section{Introduction}

Understanding the relationship between the overall macroscopic
performance of a collective, and its design at the microscopic level,
is a high priority for the collectives field (see, for example, Ref.
\cite{david2000}).  Typically such performance can be measured in
terms of a macroscopic variable of the system which fluctuates in
time: for example the level of wastage of a global resource. An
example which can be seen in other Chapters of this book, is the
discussion of the fluctuations in attendance in the El Farol bar game
and related multi-agent models such as the Minority Game. In those
cases, one is particularly interested in understanding how to minimize
the typical size of such fluctuations and hence `optimize' performance
by minimizing the average wastage. In practice, however, typical
optimization schemes which focus on minimizing the variance of some
global quantity may not be the most relevant.  Biology is a wonderful
example of a non-optimal system, yet one which does a fantastic job of
avoiding catastrophic large changes. The human body, for example, has
a complex network of feedback loops set in place in order to
`survive'.  While very few of us exist at our peak level, and even
fewer of us hold world records in athletics, we do a pretty good job
of coping with everything our changing bodies and changing environment
can throw at us. Moreover, the human system can self-manage this
difficult situation for up to 100 years, an unimaginable feat for any
computer-based system. In short, we may not be optimal in the sense of
minimizing wastage, yet our system can handle sudden changes in the
environment (i.e.\ {\em exogenous} changes) and also typically manages
to resist the tendency to self-generate large unexpected ({\em
  endogenous}) changes or `system crashes'.

Large unexpected changes, or  so-called extreme events, happen infrequently,
yet
tend to dictate the long-term dynamical behaviour of real-world
systems in disciplines as diverse as biology and economics, through to
ecology and evolution. Their consequences are often catastrophic -- for
example, the sudden jamming of traffic in an information system or on a
highway; a fatal change within the human immune system; so-called punctuated
equilibria in evolution leading to the sudden extinction of entire species;
and crashes in a financial market~\cite{bak98,sornette01,ormerod01}.  The
ability to generate large internal,
endogenous changes is a defining characteristic of complex
systems, and arguably of Nature and Life itself since it leads to evolution
through innovation. Thinking through to the management of real-world complex
systems and collectives, and in particular the risk associated with such
catastrophic changes, one wonders whether these large events could
eventually
be controlled, or even avoided altogether?

Here we take the first few steps in the direction of `risk-management' in
collectives. We consider the specific problem of a complex adaptive
multi-agent population, competing for some limited resource. The system is a
variant of Arthur's famous El Farol bar model~\cite{arthur94}, and exhibits
large,  self-generated  changes. We show that information about the large
endogenous changes becomes encoded in the system ahead of the large change
itself. The implication is that with a reasonable amount of information, the
large change would cease to be a `surprise'. The present work therefore has
relevance to both the forward and reverse problem in collectives. In
`pre-design' collectives -- i.e.\ collectives in which the individual
components can be tailor-made to have certain fixed properties -- such
knowledge could help in the design of `safe' agents whose collective
behavior is such that large changes are avoided altogether. In `post-design'
collectives -- i.e.\ collectives for which a designer has no control over the
properties of the individual components -- one could nevertheless hope to
control such large changes in order to minimize the potential damage. In
particular, one could imagine some form of soft monitoring, whereby an
external regulator monitored the system output, detected certain precursors
suggesting a dangerous build-up to a large change, then intervened to divert
it. Going further, we have recently shown~\cite{johnson02a} that such
systems can actually be `vaccinated' ahead of such large changes thereby
affording the system a degree of temporary immunity.

The science of complex systems, as befits its name, lacks a simple
definition~\cite{science99,nature01}.  It has been used to refer to systems
that are intermediate between perfect order and
perfect disorder; or even as a simple restatement of the clich\'{e}
that the behaviour of some systems as a whole can be more than the sum
of their parts.  Formal definitions of complexity do exist in the
computational and
information sciences, but apply to specific systems~\cite{boffetta01}.
 Here we focus on systems where both the properties of the individual
components and the nature of their interactions are reasonably well
understood.  The constituents themselves can be rather simple, and the
relation between any two may also be well understood, yet the
collective behavior of the ensemble still manages to escape any simple
explanation.  A simple example which captures the idea of (co-)evolution and
bounded rationality is Arthur's El Farol bar
problem~\cite{arthur94}. At its most basic level, this is an example of a
game involving a
fixed population of players (agents) $N_{tot}$ competing for a limited
resource
$L$.  No more than $L<N_{tot}$ agents can win at each timestep.  The
agents are adaptive and try to predict the next winning outcome,
which is determined only by their own choices.  Each agent is equipped
with a number of strategies from which to choose, and can do so
adaptively based on its present environment.  Recent experiments
confirm that humans indeed possess such different strategies, despite
being faced with the same history~\cite{heath02}.
In short,  the El Farol bar problem contains the following key `complex'
elements:
\begin{itemize}
\item A population of heterogeneous agents with bounded rationality.
\item Agents who use history-based strategies upon which they base their
actions.
\item A method of aggregating agent's behaviour into a global
  outcome, which introduces feedback into the system.
\end{itemize}

Our study focuses on a generic multi-agent system, based on the El Farol bar
problem, which
has already been shown to reproduce statistical and dynamical features
similar to those of a
real-world complex adaptive system: a financial market. (See
www.unifr.ch/econophysics for a detailed account of the extent to which
financial markets represent complex systems, and their associated
statistical and dynamical properties).
Our model also exhibits the crucial feature of seemingly spontaneous large
changes of variable duration~\cite{johnson00,lamper02}.
Although our study is therefore specific to a particular model, we believe
that {\em any} multi-agent model which shares the above elements would
benefit from the
techniques described in this chapter.
Recently the concept of an \emph{agent} has become increasingly
important in both artificial intelligence and mainstream computer
science~\cite{jennings01}.  In contrast to the agents considered within
El Farol-like complex systems as in this Chapter, the agents in these other
fields are typically more
sophisticated and are often utilized to perform specific tasks in a
wide range of different areas, e.g.\ e-commerce, classification,
information retrieval and management of
networks~\cite{agent1,ukmas01}.  By studying complex systems involving
more basic agents, but which retain the ability to react to their
present environment, we aim to develop approaches that
may be useful in more
complicated collectives comprising richer components.

The Chapter is arranged as follows. In \S\ref{sec:GCBG} we introduce
our generic complex system and in \S\ref{sec:large} demonstrate its
ability to generate time-series which include occasional large
movements.  In \S\ref{sec:xtremevents} we investigate the game
dynamics during these large changes and determine what is occurring
microscopically within the model system.  In \S\ref{sec:Nweight} we
introduce a method of understanding when a large change is possible
within the system, and obtain approximate expressions for its duration
in \S\ref{sec:crashlength}.
This leads to an understanding of how such large changes may
be avoided.

\section{The Grand Canonical Bar Game} \label{sec:GCBG}
We consider a generic complex system in which a population of
$N_{tot}$ heterogeneous agents with limited capabilities and
information repeatedly compete for a limited global resource.  The
model is a generalization of Arthur's bar model where agents will only
take part, or play, if they are sufficiently confident of the
strategies they hold.  A strategy is a forecasting rule that generates
a prediction of the next winning outcome based on knowledge of the
recent history of winning choices.
This is the Grand Canonical Bar Game
(GCBG),\footnote{Grand Canonical is a term used in statistical physics
  to describe a system with a variable number of particles.}
It captures some of the key behavioural phenomena
that are important in collectives/complex systems; those of
competition, frustration and adaptability.  It is also a
`minimal' system of only few parameters.

\begin{figure} [tb]
\centering
\includegraphics[scale=0.7]{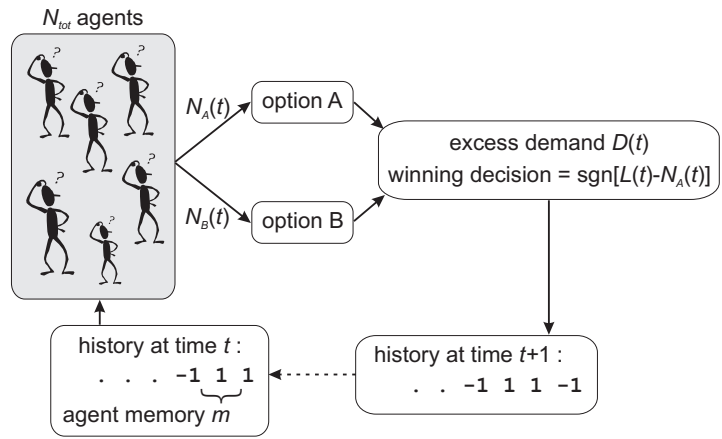}
\caption{Schematic of game structure.}
\label{fig:game}
\end{figure}

The GCBG comprises a number of agents $N_{tot}$ who repeatedly decide
whether to enter a game where the choices are option A or B.
Because of the limited global resource, only agents who are
sufficiently confident of winning will participate at each timestep.
The outcome at each timestep represents the winning decision, A or B.
The agents are adaptive in their strategy choices, but not
evolutionary; there is no discovery of new strategies by the agents.
A maximum of $L(t)$ agents can win at each timestep.  Changing $L(t)$
affects the system's quasi-equilibrium; hence $L(t)$ can be used to
mimic the changing external environment. In the limit that $L(t)$ is
time-independent, and all agents are forced to play at each timestep,
our model reverts to Arthur's El Farol bar model.
A schematic of the game structure is shown in
Figure~\ref{fig:game}, clearly indicating the feedback present within
the system. The method of encoding the outcomes in the game via a binary
alphabet, and the associated strategy space, were introduced by Challet and
Zhang~\cite{challet97}. However the GCBG differs from the basic Minority
Game in its use of an external resource level, and a variable number of
participating agents per timestep as a result of the finite confidence
level.
 The agents are of limited yet similar capabilities. Each agent is
assigned a `brain-size' $m$; this is the length of the past history
bit-string that an agent can use when making its next decision.  A
common bit-string of the $m$ most recent outcomes is made available to
the agents at each timestep.  This is the only information they can
use to decide which option to choose in subsequent timesteps.  Each
agent randomly picks $q$ strategies at the beginning of the game, with
repetitions allowed.  A strategy uses information from the historical
record of winning options to generate a prediction~\cite{challet97}.  After
each turn,
the agent assigns one (virtual) point to each of his strategies which
would have predicted the correct outcome, and minus one for an incorrect
prediction. The resulting time-series appears `random' yet is
non-Markovian, with subtle temporal correlations which put it beyond
any random-walk based description. Multi-agent games such as the present
GCBG may be simulated on a computer, but can also be
expressed in analytic form.  In the next section we introduce some
notation used in the remainder of this chapter.

\subsection{Notation}
A subset of the agent population, who
are sufficiently confident of winning, are active at each
timestep.
At each timestep $t$ we denote the number of agents choosing
option A as $N_{A}(t)$ and the number choosing B as $N_{B}(t)$.
If $L(t)-N_{A}(t)>0$ the winning decision is A and vice-versa.
The winning decision is thus given by
\begin{eqnarray*}
w(t)&=&\mbox{sgn}\left[L(t)-N_{A}(t)\right],
\end{eqnarray*}
where $\mbox{sgn}[x]$ is the sign function defined by
\begin{equation*}
\mbox{sgn}[x] = \begin{cases} -1 \quad &\text{for $x<0$}, \\ 0 &
  \text{for $x=0$}, \\ 1 & \text{for $x>0$}.\end{cases}
\end{equation*}
We denote by $w(t)$ the winning option at time $t$, where a value of
$1\Rightarrow$ option A, and $-1\Rightarrow$ option B.  We frequently
have to represent the two possible choices, A or B, in numerical form,
and we use the encoding that $1$ always implies option A and $-1$
implies option B.  If $w(t)=0$, indicating no clear winning option, this
value is
replaced with a random coin toss.

\begin{figure} [tb]
\centering
\includegraphics[scale=0.7]{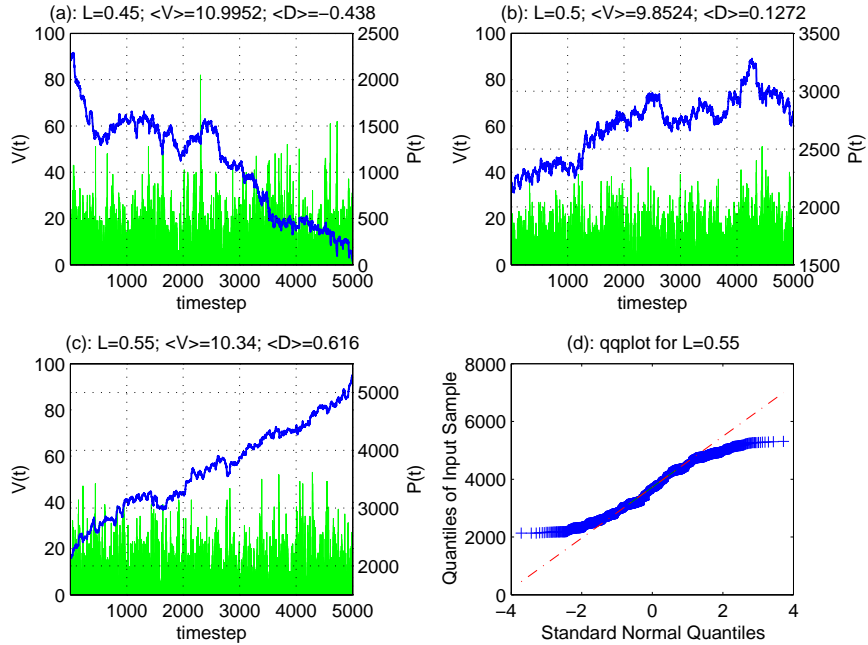}
\caption{
  Typical time-evolution of the complex adaptive system for differing
  values of the global resource level: (a) $L=0.45 V(t)$, (b) $L=0.5 V(t)$,
and
  (c) $L=0.55 V(t)$. In each case, the thick line $P(t)$ represents the
  cumulative excess demand.  For $L$ greater (less) than $0.5N_{tot}$,
$P(t)$
  has an upward (downward) trend.  (d) Corresponding qq-plot for
  $L=0.55 V(t)$ series illustrating the presence of power-law like fat
  tails. The result for a normal distribution would be a diagonal
  line, as indicated.}
\label{fig:GCMGts}
\end{figure}

The `excess demand' $D(t)=N_{A}(t)-N_{B}(t)$ (which mimics
price-change in a market) and number $V(t)=N_{A}(t)+N_{B}(t)$ of
active agents (which mimics volume) represent output
variables. These two quantities fluctuate with time, and can be
combined to construct other global quantities of interest for the
complex system studied.  We define the cumulative excess demand as
$P(t)$.  In the context of a financial market, this can be regarded as
a pseudo-price.  Typically we use the example of a financial market,
but the excess demand can be interpreteted in many different
circumstances, e.g.\ as a measure of resource utilization within a
system.  If we define $L(t)=\phi V(t)$, where $0\leq\phi\leq 1$, then
only a fraction $\phi$ of the active population can win.  By varying
the value of $\phi$, it is possible to change the equilibrium value of
$D(t)$, see Figure~\ref{fig:GCMGts}.

The only global information available to the agents is a common
bit-string `memory' of the $m$ most recent outcomes.
Consider $m=2$; the $P=2^{m}=4$ possible history bit-strings are $AA$,
$AB$, $BA$ and $BB$, with the right-most letter representing the
winning choice at the last timestep.  The history, or equivalently, the
global information available to the agents, can be represented in
decimal form $\mu\in\{0,1,\ldots,P-1\}$:
\begin{eqnarray*}
\mu(t)&=&\sum_{i=1}^{m}2^{i-1} \left[w(t-i)+1\right].
\end{eqnarray*}
The global information $\mu(t)$ is updated by dropping the first bit
and concatenating the latest outcome to the end of the string.

%
At the beginning of the game, each agent randomly picks $q$ ($>1$)
strategies, making the agents heterogeneous in their strategy
sets.\footnote{Agents are only adaptive if they have more than one
  strategy to play with; for $q=1$ the game has a trivial periodic
structure.} This initial strategy assignment is fixed from
the outset of each simulation and provides a systematic disorder which
is built into each run.  This is referred to as the \emph{quenched
  disorder} present in the game.  An agent decides which option to
choose at a given time step based on the prediction of a strategy, which
consists of a response, $a^{\mu(t)}\in\{-1,1\}$ to the global
information $\mu(t)$.  For its current play, an agent chooses its
strategy that would have performed best over the history of the game
until that time, i.e.\ has the most virtual points.

Agents have a time horizon $T$ over which virtual points are
collected and a threshold probability level $\tau$ which mimics a
`confidence'.  Only strategies having $\geq r$ points are used, where
\begin{eqnarray*}
r&=&T(2\tau-1).
\end{eqnarray*}
We call these \emph{active} strategies.  Agents with no active
strategies within their individual set of $q$ strategies do not play
at that timestep and become temporarily inactive. Agents with one or
more active strategies play the one with the highest virtual point
score; any ties between active strategies are resolved using a
coin-toss.
If an agent's threshold to play is low, we would expect the agent to
play a large proportion of the time as their best strategy will have
invariably scored higher than this threshold.  Conversely, for high
$\tau$, the agent will hardly play at all.
The coin-tosses
used to resolve ties in decisions (i.e.\ $N_{A}=N_{B}$) and
active-strategy scores, inject stochasticity into the game's
evolution.  The implementation of the strategies is discussed in
\S\ref{sec:sspace}.

After each turn, agents
update the scores of their strategies with the
reward function
\begin{eqnarray} \label{eq:MGreward}
\chi[N_{A}(t),L(t)]=\mbox{sgn}\left[N_{A}(t)-L(t)\right],
\end{eqnarray}
namely $+1$ for
choosing the correct/winning outcome, and $-1$
for choosing the incorrect outcome.
The virtual points for strategy $R$ are updated via
\begin{eqnarray*}
S_{R}(t) &=& \sum_{i=t-T}^{t-1} a_{R}^{\mu(i)} \ \chi \Big[N_{A}(i),
L(i)\Big],
\end{eqnarray*}
where $a_{R}^{\mu(t)}$ is the response of strategy $R$ to
global information $\mu(t)$, and the summation is taken over a rolling
window of fixed length
$T$.\footnote{It is also possible to construct an exponentially
  weighted window of characteristic length $T$, with a decay parameter
$\lambda=1-1/T$.
The strategy score updating equation (\ref{eq:sscore}) becomes
\begin{eqnarray*} \label{eq:sscore_exp}
\mathbf{S}(t+1)&=&\mathbf{a}^{\mu(t)}w(t)+\lambda\mathbf{S}(t).
\end{eqnarray*}
This has the advantage of removing the hard cut off from the
rolling window (a Fourier transform of the demand can show the effect of
this periodicity in the
time series), and is a rapid recursive calculation.  But strategy
scores are no longer integers, and the probability of a strategy
score tie is now very low.  Since one source of stochasticity has
now been effectively removed, there must be occasional periods of
inactivity to inject stochasticity which helps to prevent group,
or cyclic, behaviour when the game traces out a deterministic
path.}  To start the simulation, we set the initial strategy scores to be
zero: $S_{R}(0)=0$.
Because of the feedback in the game, any particular strategy's
success is short-lived. If all the agents begin to use similar
strategies, and hence make the same decision, such a strategy
ceases to be profitable and is subsequently dropped.  This encourages
heterogeneity amongst the agent population.

The demand $D(t)$ and volume $V(t)$, which can be identified as the output
from
the model system, are given by
\begin{subequations}
\begin{eqnarray} \label{eq:gcbgdemand}
D(t)&=&\mathbf{n}(t) \cdot \left( \mathbf{a^{\mu(t)}} {\cal
H}\Big[\mathbf{S}(t)-r \Big] \right)
\ = \ \sum_{R=1}^{Q}n_{R}(t)a^{\mu(t)}_{R}{\cal H}\Big[S_{R}(t)-r
\Big], \\
V(t)&=&\mathbf{n}(t) \cdot {\cal H}\Big[\mathbf{S}(t)-r \Big]
\ = \ \sum_{R=1}^{Q}n_{R}(t){\cal H}\Big[S_{R}(t)-r \Big].
\end{eqnarray}
\end{subequations}
where $n_{R}(t)$ represents the number of agents playing strategy $R$
at timestep $t$.

The demand $D(t)$ is made up of two groups of agents at each timestep:
$D_{D}(t)$ agents who act in a deterministic manner, i.e.\ do not
require a coin toss to decide which choice to make - this is because
they either have one strategy that is better than their others, or
because their highest-scoring strategies are tied but give the same
response to the history $\mu(t)$, and $D_{U}(t)$ agents
that act in an `undecided' way, i.e.\ they require the toss of a coin
to decide which choice to make - this is because they have two (or
more) highest-scoring tied strategies which give different responses
at that turn of the game. Inactive agents do not contribute to the demand.
Hence we can rewrite (\ref{eq:gcbgdemand}) as
\begin{eqnarray} \label{eq:MGDt}
D(t)&=&D_{D}(t)+D_{U}(t).
\end{eqnarray}
Without the stochastic influence of the undecided
agents the game will tend to exhibit group, or cyclic behaviour, where
the game eventually traces out a deterministic path.
The period of this cyclic behaviour
is dependent on the quenched disorder present in the simulation, and
could be very long.

The number of agents holding a particular combination of strategies
can also be expressed as a $q$-dimensional tensor
$\boldsymbol{\Omega}$~\cite{johnson01c}, where the entry
$\Omega_{R1,R2,\ldots}$ represents the number of agents holding strategies
$\{R1,R2, \ldots \}$.  This quenched disorder is fixed at the
beginning of the game.  It is useful to construct a symmetric
configuration $\boldsymbol{\Psi}$ in the sense that
$\Psi_{R1,R2,\ldots}=\Psi_{p\{R1,R2,\ldots\}}$ where
$p\{R1,R2,\ldots\}$ is any permutation of the strategies $R1,R2,\ldots$;
for $q=2$ we let
$\boldsymbol{\Psi}=\frac{1}{2}(\boldsymbol{\Omega}+\boldsymbol{\Omega}^{T})$
.
Elements $\Psi_{R,R^{\prime}}$ enumerate
the number of agents holding both strategy $R$ and $R^{\prime}$.
We focus on $q=2$ strategies per agent, although the
formalism can be generalised. At timestep $t$, $D_{D}(t)$
can now be expressed as
\begin{eqnarray*}
D_{D}(t)&=&\sum_{R=1}^{Q}a_{R}^{\mu(t)}{\cal
H}[S_{R}(t)-r]\sum_{R^{\prime}=1}
^{Q}\left(  1+\mbox{sgn}\left[  S_{R}(t)-S_{R^{\prime}}(t)\right]
\right)  \Psi_{R,R^{\prime}}.
\end{eqnarray*}
The number of undecided agents $N_{U}$ is given by
\begin{eqnarray*}
N_{U}(t)&=&\sum_{R,R^{\prime}}{\cal H}[S_{R}(t)-r]
\delta(S_{R}(t)-S_{R^{\prime}}(t))[1-\delta(a^{\mu(t)}_{R}-a^{\mu(t)}_{R^{\prime}})]\Psi_{R,R^{\prime}}
\end{eqnarray*}
and hence the demand of the undecided agents $D_{U}(t)$ is distributed
binomially:
\begin{eqnarray*}
D_{U}(t)&=&2 \ \mbox{Bin}\left(N_{U}(t),\frac{1}{2}\right)-N_{U}(t)
\end{eqnarray*}
where $\mbox{Bin}(n,p)$ is a sample from a binomial distribution of
$n$ trials with probability of success $p$.


\subsection{The strategy space} \label{sec:sspace}
The strategy space analysis in the Section was inspired by the work of
Challet and Zhang~\cite{challet97,challet98}.
A strategy consists of a response, $a^{\mu }\in\{-1,1\}$ to
each possible bit-string $\mu$, $a^{\mu}=1\Rightarrow$ option A, and
$a^{\mu}=-1\Rightarrow$ option B.  Consider $m=2$, each strategy can
be represented by a string of $P=4$ bits $[\texttt{i j k l}]$ with
$i,j,k,l=-1$ or $+1$ corresponding to the decisions based on the
histories AA, AB, BA and BB respectively.  For example, strategy
$[\texttt{-1 -1 -1 -1}]$ corresponds to deciding to pick option B
irrespective of the $m=2$ bit-string.  $[\texttt{1 1 1 1}]$
corresponds to deciding to pick option A irrespective of the $m=2$
bitstring.  $[\texttt{1 -1 1 -1}]$ corresponds to deciding to pick
option A given the histories AA or BA, and pick option B given the
histories AB or BB.
A subset of strategies can further be classed as one of the following:
\begin{itemize}
\item \emph{anti-correlated}: for example, any two agents using the
  strategies $[\texttt{-1 -1 -1 -1}]$ and $[\texttt{1 1 1 1}]$
  would take the opposite action irrespective of the sequence of
  previous outcomes.  Hence one agent will always do the opposite of
  the other agent, and their net effect on the excess demand $D(t)$
  will be zero.
\item \emph{un-correlated}: for example, any two agents using the
  strategies $[\texttt{-1 -1 -1 -1}]$ and $[\texttt{1 -1 1 -1}]$ would take
  the opposite action for two of the four histories, and
  take the same action for the remaining two histories.  Assuming that
  the $m=2$ histories occur equally often, the actions of the two
  agents will be uncorrelated on average

\end{itemize}
A convenient measure of the distance of any two strategies is the
relative Hamming distance, defined as the number of bits that need to
be changed in going from one strategy to another.  For example, the
Hamming distance between $[\texttt{-1 -1 -1 -1}]$ and $[\texttt{1 1 1 1}]$
is 4, while the Hamming distance between $[\texttt{-1 -1 -1 -1}]$ and
$[\texttt{1 -1 1 -1}]$ is just 2.

\begin{figure} [tb]
\centering
\includegraphics[scale=0.8]{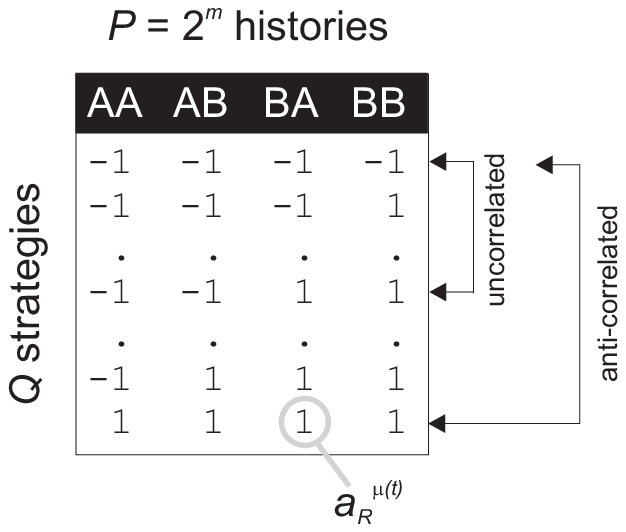}
\caption{Stylistic example of the $m=2$ strategy space.}
\label{fig:sspace}
\end{figure}

The collection of all possible strategies (and their associated
virtual points) is hereafter referred to as the \emph{strategy space},
see e.g.\ Figure~\ref{fig:sspace}.  This object can be thought of as a
common property of the game itself, being updated centrally after
each timestep.  Each individual agent monitors only a fixed subset $q$
of the $Q$ possible strategies (with the small caveat that a repeated
strategy choice is possible).  After each turn, an agent assigns one
(virtual) point to each of its strategies which would have predicted
the correct outcome.
%
The virtual points for each strategy
can be represented by the strategy score vector
$\mathbf{S}(t)$, and are given by
\begin{eqnarray} \label{eq:sscore}
\mathbf{S}(t)  &=& \sum_{i=t-T}^{t-1}\mathbf{a}^{\mu(i)} \chi \Big[N_{A}(i),
L(i)\Big]
\end{eqnarray}
In total, there are $Q=2^{P}$ possible strategies which define the
decisions in response to all possible $m$ history bit-strings.  This
is referred to as the full strategy space (FSS).
However, the principal features of the system are
reproduced in a smaller reduced strategy space (RSS) of $Q=2^{m+1}$
strategies wherein any two strategies are either un-correlated or
anti-correlated~\cite{challet98},  i.e.\ separated by a Hamming distance of
either $2^{m}$ or $2^{m-1}$.  The full and reduced strategy space
$\mathbf{a}$ for $m=2$ has been reproduced in Figure~\ref{tab:sspace}.
Each row represents a strategy, each column is assigned to a
particular history, giving the strategy space a dimension of $P \times Q$.
The prediction of strategy $R$ to information $\mu$ is $a^{\mu}_{R}$
and corresponds to the $(R,\mu)$
element of $\mathbf{a}$.  The ordering
of the rows in unimportant.

\begin{figure}[ht]
\begin{center}
\begin{gather*}
\mathbf{a} = \begin{pmatrix}
-1 & -1 & -1 & -1 \\ -1 & -1 & -1 & 1 \\
-1 & -1 & 1 & -1 \\ -1 & -1 & 1 & 1 \\
-1 & 1 & -1 & -1 \\ -1 & 1 & -1 & 1 \\
-1 & 1 & 1 & -1 \\ -1 & 1 & 1 & 1 \\
1 & -1 & -1 & -1 \\ 1 & -1 & -1 & 1 \\
1 & -1 & 1 & -1 \\ 1 & -1 & 1 & 1 \\
1 & 1 & -1 & -1 \\ 1 & 1 & -1 & 1 \\
1 & 1 & 1 & -1 \\ 1 & 1 & 1 & 1
\end{pmatrix} \qquad
\mathbf{a} = \begin{pmatrix}
-1 & -1 & -1 & -1 \\ -1 & 1 & -1 & 1 \\
1 & 1 & -1 & -1 \\ 1 & -1 & -1 & 1 \\
1 & 1 & 1 & 1 \\ 1 & -1 & 1 & -1 \\
-1 & -1 & 1 & 1 \\ -1 & 1 & 1 & -1 \\
\end{pmatrix}
\end{gather*}
\end{center}

\caption{Example of an $m=2$ strategy space.  The full strategy space
  containing 16 strategies is reproduced on the left, with the reduced
  strategy space of 8 strategies on the right.} \label{tab:sspace}
\end{figure}

%
%

Within the strategy space, each strategy $R$ has an anticorrelated
strategy $\bar{R}$.
We note that the anticorrelated strategies are effectively
redundant, as their predictions and strategy scores can be
recovered from their anticorrelated pair:
\begin{subequations}
\begin{eqnarray} \label{eq:adred}
\mathbf{a}_{\bar{R}}^{\mu}&=&-\mathbf{a}_{R}^{\mu}, \\ \label{eq:sdred}
\mathbf{S}_{\bar{R}}(t)&=&-\mathbf{S}_{R}(t).
\end{eqnarray}
\end{subequations}
Equation (\ref{eq:adred}) is true by definition, i.e.\ a strategy and its
anticorrelated pair always give the opposite prediction.  However
(\ref{eq:sdred}) requires a symmetric scoring rule to be used, which
is satisfied by (\ref{eq:MGreward}).  Thus we can reproduce the
dynamics using a space of just $P$ strategies, in which
agents choose both a strategy and whether to agree or disagree with
its prediction.  Reducing the size of the strategy space is
advantageous as it reduces memory requirements and
increases the speed of the simulation.  For a more detailed
description of how to implement such a system, see~\cite{lamper02a}.

\section{Demonstration of large changes} \label{sec:large}
\begin{figure} [tb]
\centering
\includegraphics[scale=0.7]{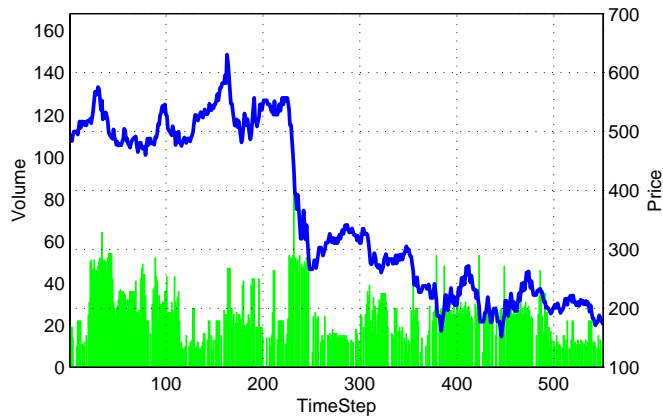}
\caption{Example of a time series exhibiting large changes.  The model
  parameters were $N_{tot}=101$, $T=60$, $\tau=0.53$, $m=3$ and $q=2$.
}
\label{fig:lchange}
\end{figure}

A ubiquitous feature of complex systems is that large changes, or
`extreme events', arise far more often than would be expected if the
individual agents acted independently.
We frequently refer to crashes, but are interested in large moves in either
direction.
With a suitable choice of
parameters the GCBG is able to
generate time series which include occasional large movements, see
e.g.\ Figure~\ref{fig:lchange}.
The game can be broadly classified into
three regimes:
\begin{enumerate}
\item The number of strategies in play is much
greater than the total available: groups of traders will play
using the same strategy and therefore crowds should dominate the
game~\cite{johnson99a}.
\item The number of strategies in play is
much less than the total available: grouping behaviour is
therefore minimal.
\item The number of strategies in play is
comparable to the total number available.
\end{enumerate}
We focus on the third regime, since this yields seemingly
random dynamics with occasional large movements.

Large changes seem to exhibit a wide range of possible durations and
magnitudes making them difficult to capture using traditional
statistical techniques based on one or two-point probability
distributions. A common feature, however, is an obvious trend (i.e.\
to the eye) in one direction over a reasonably short time window: we
use this as a working definition of a large change.
(In fact, all the large changes discussed here represent $>3\sigma$ events.)
In both our model and the
real-world system, these large changes arise more frequently than would be
expected from a random-walk model~\cite{sornette01}.

In \S\ref{sec:tail} we consider the distribution of the excess demand
created by our generic system, and perform a simple analysis to
discuss the statistics of extreme events.  To determine whether large
movements occur due to a single random event, we examine the
stochastic influence present within the model in \S\ref{sec:stoc} and
find that they arise through a global cooperation occurring over the
whole system.

\subsection{Tail estimation} \label{sec:tail}
%
Traditional parametric statistical methods are
ill-suited for dealing with extreme events,
which have little historical data associated with them.
Provided the distribution has a finite variance and the
increments are independent, the Central Limit Theorem will apply
near the centre of the distribution, but does not tell us
anything about the tails~\cite{bouchaud00}.\footnote{The speed at which
the distribution will converge to a Gaussian is given by the
Berry-Esseen theorem; a finite second moment ensures Gaussian
behaviour, and the speed of convergence is controlled by the size
of the third moment of $\vert x \vert$~\cite{shlesinger95}.}
In recent years extreme value theory (EVT), a branch of probability
theory which focuses explicitly on extreme outcomes, has received
increasing attention~\cite{embrechts99}.
EVT considers
the distribution of extreme returns rather than the distribution of
all returns.  It can be shown that the limiting distribution of
extreme returns observed over a long time-period is largely
independent of the distribution of returns itself.
The upper tail of a fat-tailed cumulative distribution function $F$ behaves
asymptotically like the tail of the Pareto distribution
given by $1-F(x) \approx cx^{-\alpha}$, for $c>0$, $\alpha>0$ and
$x\geq C$, where $C$ is a threshold above which the assumed algebraic
form is valid, and $c$ is a normalising constant.  The tail index
$\alpha$ determines the heaviness of the tail of a distribution and
plays a key role in tail-related risk measures, representing the
maximal order of finite moments.  Only the first $k$ moments, where
$k<\alpha$, are bounded.  The greater the tail index, the `fatter' the
tail, and the greater the incidence of extreme events.

\begin{figure} [tb]
\centering
\includegraphics[scale=0.7]{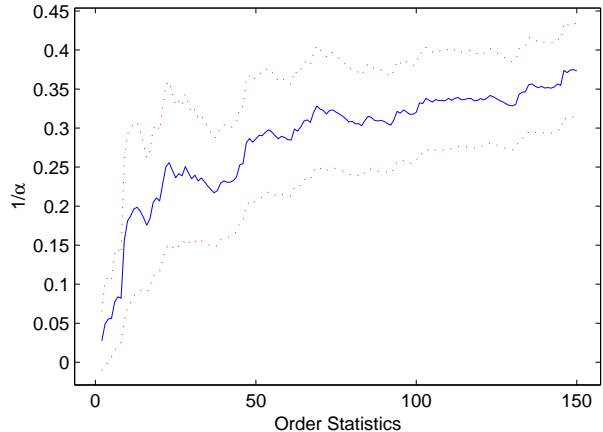}
\caption{Hill-plot with a 0.95 confidence interval.  The estimated
  value of the shape parameter $\xi$ is plotted against the number of
  upper order statistics $m$ used in the estimation.  A value of $\xi$
  is selected from the plot where the shape parameter is fairly
  stable}
\label{fig:hillplotMG}
\end{figure}

An important issue in the study of fat-tailed distributions is the
estimation of the tail index $\alpha$.
There are a number of methods to estimate the tail index, some using
asymptotic results from EVT, from which values can
be estimated using maximum likelihood techniques.
However the Hill estimator is commonly used,
as it is suitable for tail estimation of fat-tailed
distributions and is relatively easy to implement~\cite{hill75}.
Consider a sample $X_{1}, X_{2}, \ldots, X_{n}$ of $n$
observations drawn from a stationary iid process,
and let $X_{(1)} \geq X_{(2)} \geq \ldots \geq
X_{(n)}$ be the descending order statistics.
The Hill estimator is based on the difference between the $m$th
largest observation and the average of the $m$ largest observations:
\begin{eqnarray*}
\hat{\xi}_{m}&=&\frac{1}{m}\sum_{i=1}^{m}\log X_{(i)}
- \log X_{(m)}
\end{eqnarray*}
where $\xi=1/\alpha$ is the shape parameter and $m$ is the number
of order statistics used in the tail estimation.  The appropriate
choice of value $m$ is a non-trivial problem, since this requires us
to decide where the tail begins.
There is a tradeoff between the bias and variance of the
estimator in choosing $m$.  If we choose a large value of $m$, the
number of order statistics used increases and the variance of the
estimator will decrease.  However, choosing a high $m$ also introduces
some observations from the centre of the distribution and the
estimation becomes biased.  But if it is too small, the estimate will
be based on a just a few of the largest observations and the estimator
will lack precision.
Several methods for the determination of an optimal sample
fraction for the Hill estimator have been proposed~\cite{daniel01,drees98}.
In Figure~\ref{fig:hillplotMG} a Hill plot is constructed from the
simulation data used to create Fig.~\ref{fig:lchange}.
A threshold is selected from the plot where the shape parameter is
fairly stable, giving an estimate of the tail index $\alpha \approx
3.7$.
To obtain the most accurate tail estimate, a combination of several
techniques should be considered.

\subsection{The effect of stochasticity} \label{sec:stoc}
To investigate the stochastic influence due to the effect of coin tosses,
we have studied their occurrence during the simulation.
We define two types of coin toss: type (i), which occurs when an agent
has a tie between active strategy options which predict differing
outcomes, and type (ii), which
occurs when the number of agents choosing option A is equal to the
number choosing option B.   The main conclusion is that there is no
single-coin toss that immediately causes a large change within the system.
%
%
and
%
The large movements observed as
 arising endogenously in the system, result
 from the organisation of temporal and spatial (i.e. strategy space)
 correlations. In common with other complex adaptive systems, this
 organisation does not arise in general from a nucleation phase diffusing
 across the system, e.g. it cannot be traced to a particular coin-toss by a
 particular agent which triggers the ``avalanche''. Rather, it results from
a
 progressive and more global cooperation occurring over the whole system via
 repetitive interactions~\cite{sornette00a}.

\section{A study of extreme events} \label{sec:xtremevents}

In this section we consider the generic complex system introduced in
\S\ref{sec:GCBG} in which a population of $N_{tot}$ heterogeneous
agents with limited capabilities and information, repeatedly compete
for a limited global resource.  We focus on extreme events which are
endogenous (i.e.\ internally-produced by the agents themselves)
and provide a microscopic understanding as to their build-up and
likely duration.

\subsection{Nodal weight decomposition} \label{sec:Nweight}
\begin{figure} [tb]
\centering
\includegraphics[scale=0.8]{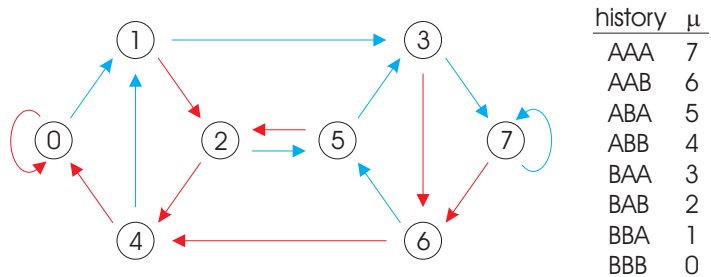}
\caption{Dynamical behaviour of the global information is described by
transitions on the de Bruijn graph. Graph for population of $m=3$\ agents.
Blue transitions represent positive demand $D$, red transitions represent
negative demand, with each transition incurring an increment to the score
vector $\mathbf{S}$.}
\label{fig:deBruijn}
\end{figure}

The dynamics of the game history can be usefully represented on a
directed de Bruijn graph~\cite{metzler02}.  This is a graph whose nodes
are sequences of symbols from some alphabet, and whose edges represent
possible transitions between these nodes.  It has numerous interesting
properties and is frequently discussed in the context of parallel
algorithms and communications networks.  This is an effective method
of representing the evolution of the system, and in
Figure~\ref{fig:deBruijn} we plot the de Bruijn graph for an $m=3$
game.

Large changes occur when connected nodes become persistent and the
game makes successive moves in the same direction. Only nodes
0 and 7 can exhibit perfect nodal persistence, where an allowed
transition can return the system to exactly the same node.  This is the
simplest
type of large change, e.g.\ $\mu(t)=0,0,0,0,\ldots$, where all successive
price changes are in the \emph{same} direction. We call this a
`fixed-node crash' (or rally).  There are many other
possibilities reflecting the wide range of forms and durations that a
large change can undertake.  For example, on the $m=3$ de Bruijn graph in
Fig.~\ref{fig:deBruijn} the cycle $\mu=0,0,1,2,4,0,\ldots$ has four
out of the five transitions producing price-changes of the same sign
(it is persistent on nodes 1, 2, 4 and antipersistent on node 0). We
call this a `cyclic-node crash'.  Stable behaviour occurs on a path
where all transitions are equally visited, e.g.\
$\mu(t)=0,0,1,3,6,5,3,7,7,6,4,1,2,5,4,\ldots$.

\begin{figure} [tbf]
\centering
\includegraphics[scale=0.7]{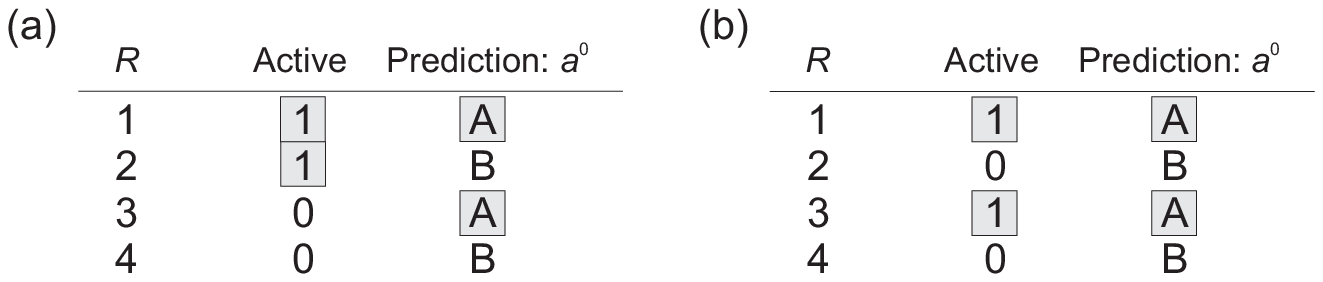}
\caption{Two different configurations of the strategy space for
  $\mu(t)=0$.  The columns represent: $R$ -- strategy index; Active -- where
  `1' represents an active strategy and `0' an inactive
  strategy, i.e.\ ${\cal H}[S_{R}(t)-r]$; Prediction -- the prediction
  of the strategy, which is given by $a^{0}_{R}$.  In (a) there is no
  consensus in the predictions of the active strategies.  The active
  strategies do not share the same pattern as $\mathbf{a}^{0}$.  In
  (b) there is a consensus of opinion. All the active strategies predict
option $A$ since the pattern
  of active strategies is similar to the pattern of strategy predictions.
}
\label{fig:nweightex}
\end{figure}

To identify moments when large changes occur, we need to
recognise when history nodes are likely to become persistent.  Whether a
node is persistant or not will depend on the action of the agents at that
timestep, which is determined by the predictions of the
strategies they hold.  If the majority of active strategies generate a
prediction of the same outcome, then the demand at that timestep is likely
to continue in
that direction.  Thus a suitable condition for a large movement is
when there is a concensus of opinion regarding the next outcome amongst
the active strategies.  This occurs when the
pattern of active strategies within the strategy space matches up with the
pattern of stratgy
predictions for a history node $\mu$, see Figure~\ref{fig:nweightex}.
The strategy predictions will depend on the global information
$\mu(t)$, and the distribution of active strategies depends on the
strategy score vector $\mathbf{S}(t)$.
%
At each timestep $\mathbf{S}(t)$ is
updated according to whether a strategy predicted the winning outcome
$w(t)$; the incremental strategy score is given by a column of the
strategy matrix $\mathbf{a}$ determined by $\mu(t)$.  In total there are $P$
orthogonal
increment vectors $\mathbf{a}^{\mu}$, one for each value of $\mu$.  We
can express the strategy score (\ref{eq:sscore}) as
\begin{eqnarray}
\mathbf{S}(t)=c_{0}\mathbf{a}^{0}+c_{1}\mathbf{a}^{1}+\ldots
+c_{P-1}\mathbf{a}^{P-1}\ =\ \sum_{j=0}^{P-1}c_{j}\mathbf{a}^{j},
\end{eqnarray}
where $c_{j}$ represents the \emph{nodal weight} for history node $\mu=j$.
The nodal weights represent the number of negative return transitions from
node $\mu$ minus the number of positive return transitions, in the time
window
$t-T\rightarrow t-1$.  The values of these nodal weights are important
in identifying periods in which large changes can occur.
If the value of a nodal weight is near zero, this implies that the
number of active strategies predicting outcome A will be similar to
the number predicting outcome B when that node is reached.  The excess
demand will then depend on the quenched disorder and the undecided agents,
but is likely to be small.  Conversely, if the value of a nodal weight
is significantly different from zero, this indicates a bias in the
strategy space, i.e.\ the same pattern is evident in the active
strategies and the strategy predictions.  When the game trajectory
hits this node, the majority of active strategies will predict
the same outcome, and there will be a consensus amongst the agents.
This is likely to lead to a large excess demand, and therefore a large
change.  Thus high absolute
nodal weight implies persistence in transitions from that node, i.e.\
persistence in $D|\mu$.

The mean of the strategy scores predicting an $A$ or $B$ at node $\mu$
are linked to their nodal weight:
\begin{subequations}
\begin{eqnarray} \label{eq:meanSp}
\mbox{E}\left[S_{R\ni a_{R}^{\mu}=1}(t)\right]&=&c_{\mu}(t), \\
\label{eq:meanSn}
\mbox{E}\left[S_{R\ni a_{R}^{\mu}=-1}(t)\right]&=&-c_{\mu}(t).
\end{eqnarray}
\end{subequations}
This symmetry occurs because each strategy has an anticorrelated
strategy present in the strategy space.

\begin{figure} [tb]
\centering
\includegraphics[scale=0.6]{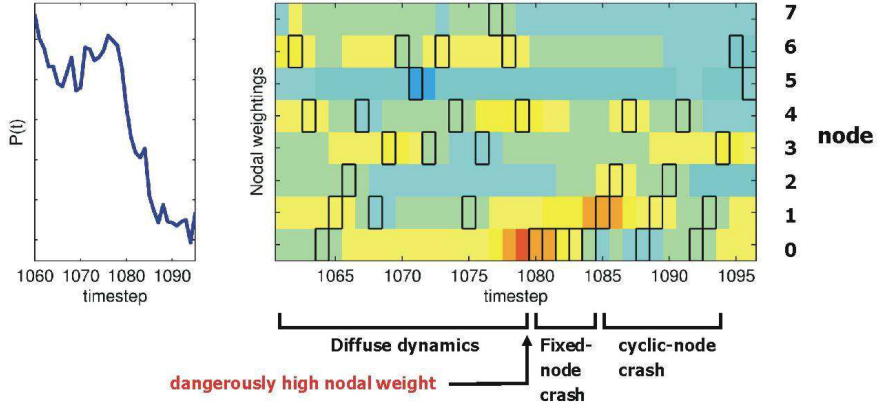}
\caption{Dynamical behaviour of complex system (e.g. price $P(t)$ in
  financial market) described by evolution of nodal weights $c_{\mu}$.
  History at each timestep indicated by black square. Large change
  preceded by abnormally high nodal weight. Large change incorporates
  fixed-node and cyclic node crashes. }
\label{fig:nodalweights}
\end{figure}

The nodal weight decomposition provides a succinct method of describing
the state of the strategy space.  We are concerned with high nodal
weights on game cycles that can become persistent.  When a nodal
weight value becomes large, this is a warning sign that the system is
in a suitable state to undergo a large change.
Figure~\ref{fig:nodalweights} illustrates a large change which starts
as a fixed-node crash, and then subsequently becomes a cyclic-node crash.

\subsection{Estimating the crash length} \label{sec:crashlength}
We are interested in the dynamics of large changes, and use a
simplified version of the system to obtain an analytic expression for
the expected crash length.  Reference~\cite{johnson02} showed that the
GCBG can be usefully described as a stochastically disturbed
deterministic system.  As stated in (\ref{eq:MGDt}), the demand can be
divided into contributions arising from agents acting on a definite
strategy $D_{D}(t)$, and that from the undecided agents $D_{U}$.
The average contribution of the undecided agents to the net demand
will be zero, i.e. $\mbox{E}[D_{U}(t)]=0$.
Averaging over our model's stochasticity in this way yields a
description of the game's deterministic dynamics.  By examining these
dynamics, we can determine when a large-change is likely.
For $q=2$ the demand function can be expressed as
\begin{eqnarray}
D(t)&=&\sum_{R=1}^{Q}a_{R}^{\mu(t)}{\cal H}[S_{R}(t)-r]\sum_{R^{\prime}=1}
^{Q}\left(  1+\mbox{sgn}\left[  S_{R}(t)-S_{R^{\prime}}(t)\right]
\right)  \Psi_{R,R^{\prime}},
\end{eqnarray}
where $\mathbf{\Psi}$ is the symmetrized strategy
allocation matrix which constitutes the quenched disorder present
during the system's evolution. The volume $V(t)$ is given by the same
expression as $D(t)$ replacing $a_{R}^{\mu(t)}$ by unity.

For the parameter ranges of interest,
the choice about whether a strategy is played by an agent is more determined
by whether that strategy's score is above the threshold, than whether it is
their highest-scoring strategy. This is because agents are
only likely to have at most one strategy whose score lies above the
threshold
for confidence levels $r\geq0$. Making the additional numerically-justified
approximation of small quenched disorder (i.e.\ the variance of the entries
in
the strategy allocation matrix $\mathbf{\Psi}$ is smaller than
their mean for the parameter range of interest~\cite{johnson02}), the demand
and volume become
\begin{subequations}
\begin{eqnarray}
D(t)&=&\sum_{R=1}^{Q}a_{R}^{\mu(t)}{\cal H}[S_{R}(t)-r]\sum_{R^{\prime}=1}
^{Q}\frac{N}{Q^{2}} \nonumber \\
&=&\frac{N}{Q}\sum_{R=1}^{Q}a_{R}^{\mu(t)}
{\cal H}\left[  S_{R}(t)-r\right] \nonumber  \\
&=&\frac{N}{Q}\sum_{R=1}^{Q}a_{R}^{\mu(t)}
\frac{1}{2}\Big(1+ \mbox{sgn}\left[  S_{R}(t)-r\right]\Big) \nonumber  \\
 &=&  \frac{N}{2Q}\sum_{R=1}^{Q}a_{R}^{\mu(t)}
\mbox{sgn}\left[  S_{R}(t)-r\right]  ,\label{eq:sqddemand}\\
V(t)  & =& \frac{N}{Q}\sum_{R=1}^{Q}{\cal H}
\left[  S_{R}(t)-r\right] \ = \
\frac{N}{2}+\frac{N}{2Q}\sum_{R=1}^{Q}\mbox{sgn}
\left[  S_{R}(t)-r\right]  .\label{eq:sqdvolume}
\end{eqnarray}
\end{subequations}

Let us suppose persistence on node $\mu=0$ starts at time $t_{0}$. How long
will the resulting large change last? To answer this, we decompose
(\ref{eq:sqddemand}) into strategies which predict option $A$ at
$\mu=0$, and those that predict $B$. We first consider the particular
case where the node $\mu=0$ was not visited during the previous
$T$ timesteps, hence the loss of score increment from time-step $t-T$
will not affect $\mathbf{S}(t)$ on average. At any later time
$t_{0}+\tau$ during the large change, (\ref{eq:sqddemand}) and
(\ref{eq:sqdvolume}) are given by
\begin{eqnarray}
D(t_{0}+\tau) &=& -\frac{N}{2Q}\left\{  \sum_{R\ni a_{R}^{\mu}%
=-1}\operatorname*{sgn}\left[  S_{R}(t_{0})-r-\tau\right]  -\sum_{R\ni
a_{R}^{\mu}=1}\operatorname*{sgn}\left[  S_{R}(t_{0})-r+\tau\right]
\right\}
,\label{eq:sqddemand2}\\
V(t_{0}+\tau) &=& \frac{N}{2}+\frac{N}{2Q}\left\{  \sum_{R\ni a_{R}%
^{\mu}=-1}\operatorname*{sgn}\left[  S_{R}(t_{0})-r-\tau\right]  +\sum_{R\ni
a_{R}^{\mu}=1}\operatorname*{sgn}\left[  S_{R}(t_{0})-r+\tau\right]
\right\}
.\nonumber
\end{eqnarray}
The magnitude of the demand $|D(t_{0}+\tau)|$ decreases as the
persistence time $\tau$ increases, and the large change will end at
time $t_{0}+\tau_{c}$ when the right-hand side of
(\ref{eq:sqddemand2}) becomes zero.  We denote the persistence
time or `crash-length' by $\tau_{c}$.  It is easy to obtain an upper limit
for the duration of a large change by determining when the demand $D$
changes sign.  This occurs when the mean of the scores of the strategies
predicting $A$ returns to 0, i.e.\
\begin{eqnarray*}
\tau_{c}&=&\mbox{E}\left[S_{R\ni a_{R}^{\mu}=-1}(t_{0})\right] \\
&=&-c_{0}(t_{0})
\end{eqnarray*}
from (\ref{eq:meanSn}).  Thus the crash length will depend on the score
difference
between strategies predicting option $A$ and those predicting $B$.
In
the more general case, where the node $\mu=0$ was visited during
the previous $T$ timesteps, $\tau_{c}$ is given by the largest $\tau$
value which satisfies
\begin{eqnarray}
\tau&=&-\left(  c_{0}(t_{0})+\sum_{\{t^{\prime}\}}\operatorname*{sgn}\left[
D(t^{\prime})\right]  \right)
\end{eqnarray}
where $\{t^{\prime}\}\ni(\mu( t^{\prime})=0\cap t_{0}-T\leq t^{\prime
}\leq t_{0}-T+\tau)$. The summation accounts for any $\mu=0$
transitions in the period $t_{0}-T$ to $t_{0}-T+\tau$.

\begin{figure} [tb]
\centering
\includegraphics[scale=0.6]{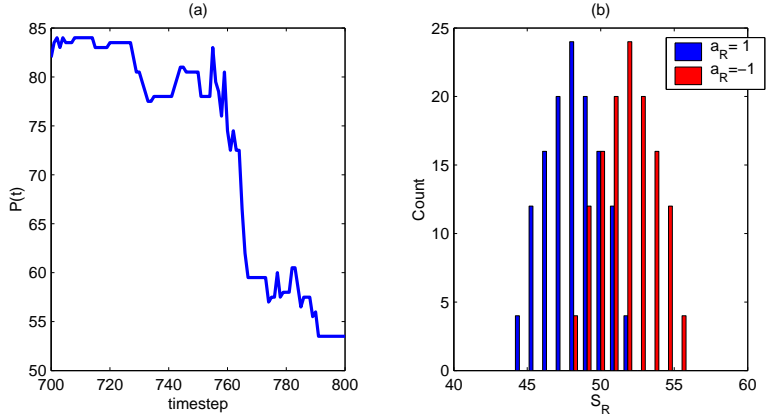}
\caption{Empirical strategy score distribution.  (b) is the strategy
  score distribution at timestep 760 from the series depicted in (a).}
\label{fig:sscore}
\end{figure}

\begin{figure} [tb]
\centering
\includegraphics[scale=0.6]{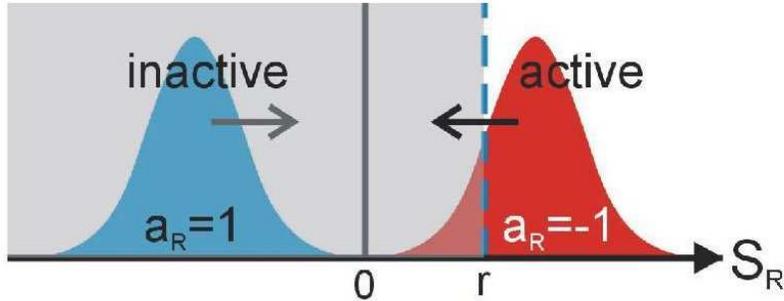}
\caption{Schematic representation of strategy score distribution prior to
 crash. Arrows indicate subsequent motion during crash period.}
\label{fig:crash3}
\end{figure}

\begin{figure} [tb]
\centering
\includegraphics[scale=0.8]{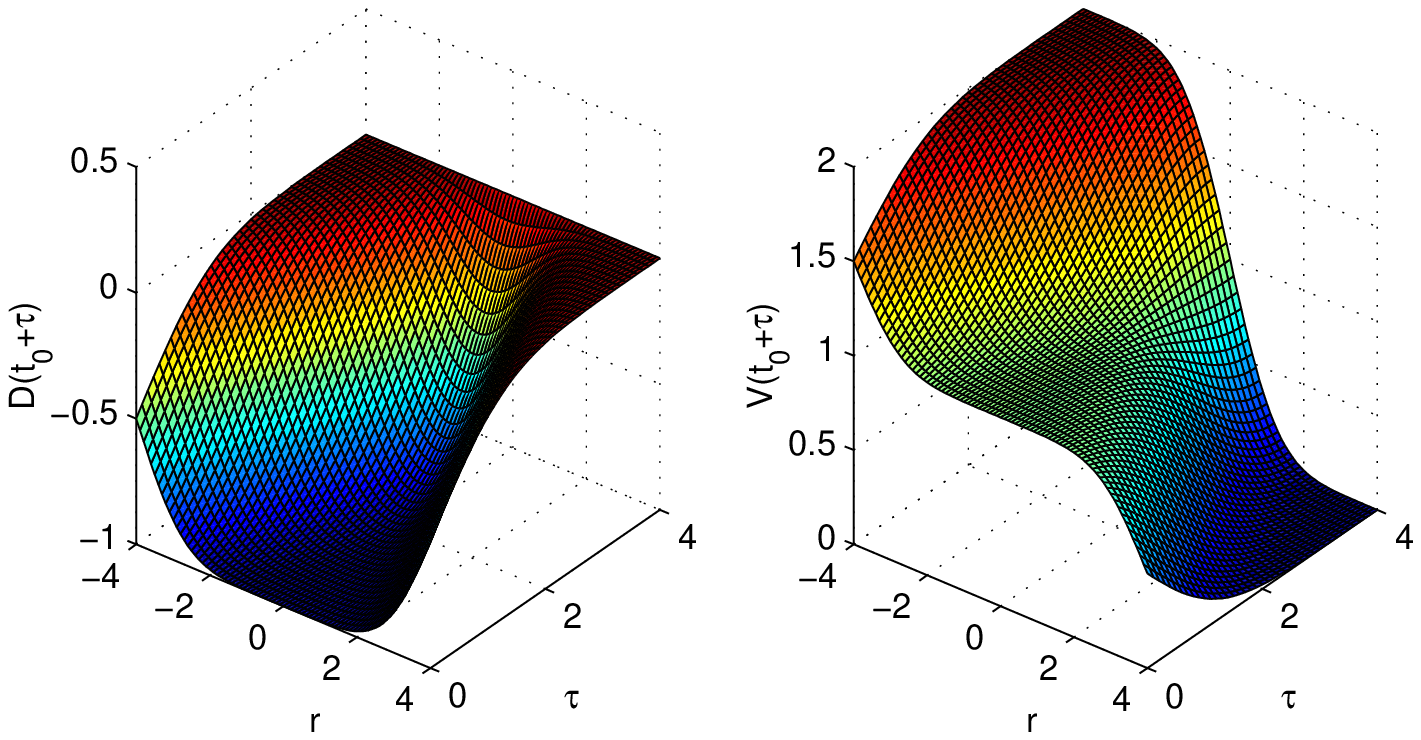}
\caption{Plots of
 expected demand $D(t)$ and volume $V(t)$ during crash period showing range
of
 different possible behaviour as system parameters are varied.}
\label{fig:crash4}
\end{figure}

To obtain an improved estimate of $\tau_{c}$, and investigate the
behaviour of the demand and volume, it is necessary to
assume a distribution for the strategy scores.  In
Figure~\ref{fig:sscore} we plot an example of the strategy score
distribution prior
to a large change.
We assume that the scores have a near-Normal
distribution, i.e.\ $S_{R\ni a_{R}^{\mu}=-1}(t_{0})\sim\operatorname*{N}%
[-c_{0}(t_{0}),\sigma]$ as shown Fig.~\ref{fig:crash3}.\footnote{At
  some points during the simulation the distribution can be more
  complex, e.g.\ multimodal behaviour.}
Consequently, prior to a large change, the score distribution tends to
split into two halves.
Substitution into (\ref{eq:sqddemand2}) gives the expected demand and volume
during the
crash:
\begin{eqnarray*}
D(t_{0}+\tau) &  \propto &
\Phi\left[\frac{c_{0}(t_{0})-r+\tau}{\sigma}\right] -
\Phi\left[\frac{-c_{0}(t_{0})-r-\tau}{\sigma}\right],  \\
V(t_{0}+\tau) &  \propto &
\Phi\left[\frac{c_{0}(t_{0})-r+\tau}{\sigma}\right] +
\Phi\left[\frac{-c_{0}(t_{0})-r-\tau}{\sigma}\right].
\end{eqnarray*}
These forms are illustrated in Figure~\ref{fig:crash4}. As the spread
in the strategy score distribution is increased, the dependence of $D$
and $V$ on the parameters $\tau$ and $r$ becomes weaker and the
surfaces flatten out leading to a smoother drawdown, as opposed to a
sudden severe crash.  In the limit $\sigma \rightarrow 0$, the
crash-length $\tau_{c}=-c_{0}(t_{0})-\vert r \vert$.  As the parameters
$c_{0},\sigma,r$ are varied, it can be seen that the behaviour of the
demand and volume during the large change can exhibit markedly
different qualitative forms.

\subsection{Repeated occurence of large changes}
\begin{figure} [bt]
\centering
\includegraphics[scale=0.6]{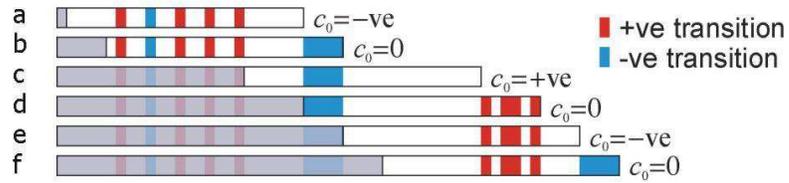}
\caption{Representation of how large changes can recur due to finite memory
of agents. Grey area shows history period outside agents' memory. Example
shows recurring fixed-node crash at node $\mu=0$.}
\label{fig:crash5}
\end{figure}

We now turn to the important practical question of whether history
will repeat itself, i.e.\ given that a large change has recently
happened, is it likely to happen again? If so, is it likely to be even
bigger? Suppose the system has built up a negative nodal weight for
$\mu=0$ at some point in the game (see Fig.~\ref{fig:crash5}a).  It
then hits node $\mu=0$ at time $t_{0}$ producing a large change
(Fig.~\ref{fig:crash5}b). The nodal weight $c_{0}$ is hence restored
to zero (Fig.~\ref{fig:crash5}c). In this model the previous build-up
is then forgotten because of the finite $T$ score window, hence
$c_{0}$ becomes positive (Fig.~\ref{fig:crash5}d). The system then
corrects this imbalance (Fig.~\ref{fig:crash5}e), restoring $c_{0}$ to
0. The large change is then forgotten, hence $c_{0}$ becomes negative
(Fig.~\ref{fig:crash5}f). The system should therefore crash again -
however, a crash will \emph{only} re-appear if the system's trajectory
subsequently returns to node $\mu=0$. Interestingly, we find that the
\emph{disorder} in the initial distribution of strategies among agents
(i.e.\ the quenched disorder in $\mathbf{\Psi}$) can play a deciding
role in the issue of `births and revivals' of large changes since it
leads to a slight bias in the outcome, and hence the subsequent
transition, at each node.  
In certain configurations, this system may 
demonstrate repeated instabilities leading to large changes.
When $c_{\mu(t)}=0$ (see
Fig.~\ref{fig:crash5}c), it follows that
$\operatorname*{sgn}[D(t)]$ is more likely to be equal to $\operatorname*{sgn}%
\left[  \mathbf{a}^{\mu(t) }\cdot\mathbf{x}\right]  $
where $x_{R}=\sum_{R^{\prime}}\Psi_{R,R^{\prime}}$
is a strategy weight vector with $x_{R}$ corresponding to the
number of agents who hold strategy $R$~\cite{jefferies02a}. The quenched
disorder therefore provides a crucial bias for determining the future
trajectory on the de Bruijn graph when the nodal weight is small,
and hence
can decide whether a given large change recurs or simply
disappears. The quenched
disorder also provides a \emph{catalyst} for building up a very
large change~\cite{jefferies02a}.

\section{Concluding remarks}
We have addressed the issue of understanding, predicting and eventually
controlling catastrophic endogenous changes in a collective.
By utilizing information about the strategy weights within the
model, we have developed a method for determining when a large change is
likely within a generic complex system, the so-called GCBG, and obtained an
analytic expression for its expected duration.
Our work opens up the study of how a `complex-systems-manager' might
use this information to control the long-term evolution of a complex
system by introducing, or manipulating, such large
changes~\cite{jefferies02a}.
As an example, we give a quick three-step solution to prevent large
changes: (1) use the past history of outcomes to build up an estimate
of the score vector $\mbox{S}(t)$ and the nodal weights
$\{c_{\mu(t)}\}$ on the various critical nodes, such as $\mu=0$ in the
case of the fixed-node crash. (2) Monitor these weights to check for
any large build-up. (3) If such a build-up occurs, step in to prevent
the system hitting that node until the weights have decreased.  Finally we
note that it can
sometimes be beneficial to induce small changes ahead of time, in order to
avoid larger changes in the future. We call this process `immunization' and
refer to Ref.\ \cite{johnson02a} for more details.

\vspace{5mm}

We are grateful to David Wolpert, Kagan Tumer and Damien Challet
for many  useful discussions about Collectives.

\clearpage

\bibliographystyle{plain}

\end{document}